\documentclass{emulateapj}
\usepackage{times}

\newcommand{\degg}{\hbox{$^\circ$}}

\newcommand{\et}{et al.\ }
\newcommand{\xmm}{{\it XMM-Newton}}
\newcommand{\sax}{{\it BeppoSAX}}
\newcommand{\swift}{{\it Swift}}

\newcommand{\ls}{\mathrel{\hbox{\rlap{\hbox{\lower4pt\hbox{$\sim$}}}\hbox{$<$}}}}
\newcommand{\gs}{\mathrel{\hbox{\rlap{\hbox{\lower4pt\hbox{$\sim$}}}\hbox{$>$}}}}
\newcommand{\grb}{GRB~050315}

\begin{document}

\submitted{Accepted 2005 October 21}
\title{\swift\ observations of the X-ray bright \grb}
\shorttitle{\swift\ observations of \grb}
\shortauthors{Vaughan \et}
\author{ S. Vaughan\altaffilmark{1}, 
M. R. Goad\altaffilmark{1},
A. P. Beardmore\altaffilmark{1},
P. T. O'Brien\altaffilmark{1},
J. P. Osborne\altaffilmark{1},
K. L. Page\altaffilmark{1},
% ---------------------
S. D. Barthelmy\altaffilmark{2},
D. N. Burrows\altaffilmark{3},
S. Campana\altaffilmark{4},
J. K. Cannizzo\altaffilmark{2,5},
M. Capalbi\altaffilmark{6},
G. Chincarini\altaffilmark{4,7},
J. R. Cummings\altaffilmark{2,8}
G. Cusumano\altaffilmark{9},
P. Giommi\altaffilmark{6},
O. Godet\altaffilmark{1},
J. E. Hill\altaffilmark{2,10},
S. Kobayashi\altaffilmark{3},
P. Kumar\altaffilmark{11},
V. La Parola\altaffilmark{9},
A. Levan\altaffilmark{1},
V. Mangano\altaffilmark{9},
P. M{\'e}sz{\'a}ros\altaffilmark{3},
A. Moretti \altaffilmark{4},
D. C. Morris\altaffilmark{3},
J. A. Nousek\altaffilmark{3},
C. Pagani\altaffilmark{3,4},
D. M. Palmer\altaffilmark{12},
J. L. Racusin\altaffilmark{3},
P. Romano\altaffilmark{4},
G. Tagliaferri\altaffilmark{4},
B. Zhang\altaffilmark{13}\\
and N. Gehrels\altaffilmark{2}
} 
\email{sav2@star.le.ac.uk}
\altaffiltext{1}{Department of Physics \& Astronomy, University of
  Leicester, Leicester LE1 7RH, United Kingdom.} 
\altaffiltext{2}{NASA Goddard Space Flight Center, Greenbelt, MD 20771} 
\altaffiltext{3}{Department of Astronomy \& Astrophysics, Pennsylvania
  State University, 525 Davey Lab, University Park, PA 16802.}
\altaffiltext{4}{INAF-Osservatorio Astronomico di Brera, Via Bianchi
  46, 23807 Merate, Italy.} 
\altaffiltext{5}{Joint Center for Astrophysics, University of
  Maryland, Baltimore County, Baltimore, MD 21250.}
\altaffiltext{6}{ASI Science Data Center, Via Galileo Galilei, I-00044
  Frascati (Rome), Italy.} 
\altaffiltext{7}{Universit\'{a} degli studi di Milano-Bicocca,
  Dipartimento di Fisica, Piazza delle Scienze 3, I-20126 Milan,
  Italy.} 
\altaffiltext{8}{National Research Council, 2101 Constitution Avenue,
  NW, Washington, DC 20418. } 
\altaffiltext{9}{INAF-Istituto di Astrofisica Spaziale e Fisica
  Cosmica Sezione di Palermo, Via Ugo La Malfa 153, I-90146 Palermo,
  Italy.} 
\altaffiltext{10}{Universities Space Research Association, 10211 Wincopin Circle, Suite 500,
Columbia, MD, 21044-3432, USA}
\altaffiltext{11}{Department of Astronomy, University of Texas,
  Austin, TX 78712} 
\altaffiltext{12}{Los Alamos National Laboratory, P.O. Box 1663, Los
  Alamos, NM 87545.} 
\altaffiltext{13}{Department of Physics, University of Nevada, Box
  454002, Las Vegas, NV 89154-4002.} 

\begin{abstract}

This paper discusses \swift\ observations of the $\gamma$-ray burst 
\grb\ (z=1.949) from  $80$~s to $10$~days after the onset of the burst.
The X-ray light curve displayed a steep early decay ($t^{-5}$) for
$\sim 200$ s and several breaks.  However, both the prompt hard
X-ray/$\gamma$-ray emission (observed by the BAT) and the first $\sim
300$ s of
X-ray emission (observed by the XRT) can be explained by exponential
decays, with similar decay constants.   Extrapolating the BAT
light curve into the XRT band suggests the
rapidly decaying, early X-ray emission was simply a continuation of
the fading prompt emission; this strong similarity between the prompt
$\gamma$-ray and early X-ray emission may be related to the simple
temporal and spectral character of this X-ray rich GRB. The prompt
(BAT) spectrum was a steep down to $\sim 15$~keV, and appeared to continue through the XRT bandpass, implying a
low peak energy, inconsistent with the Amati relation.  Following the
initial steep decline the X-ray afterglow did not fade for $\sim
1.2\times10^4$ s,
after which time it decayed with a temporal index of $\alpha\approx0.7$,
followed by a second break at $\sim 2.5\times10^5$ s to a slope of $\alpha\sim2$. The
apparent `plateau' in the X-ray light curve, after the early rapid
decay, makes this one of the most extreme examples of the
steep-flat-steep X-ray light curves revealed by \swift.  If the second
afterglow break  is identified with a jet break then the jet opening
angle  was $\theta_0\sim5$\degg, and implying $E_{\gamma} \gs 10^{50}$~erg.

\end{abstract}


%%%%%%%%%%%%%%%%%%%%%%%%%%%%%%%%%%%%%%%%%%%%%%%%%%%%%%%%%%%%%%%%%%%%%%%%%%%%%%%

\section{Introduction}

The \swift\ $\gamma$-ray burst Explorer (Gehrels \et 2004) was
successfully launched on 2004 November 20 and is now routinely taking
observations of Gamma-ray Bursts (GRBs) and their afterglows in the
crucial minutes to hours after the burst, 
delivering insights into the nature of the prompt emission and early
afterglow phase.  New bursts are detected
by the Burst Alert Telescope (BAT; Barthelmy 2004; Barthelmy \et 2005), a coded-mask imager
with a with a  $1.4$~sr field-of-view (half-coded) sensitive 
to $15-350$~keV energies, with imaging capability over the
$15-150$~keV range.
The spacecraft is able to slew autonomously to the burst position
within a few tens of seconds.  Once on-target, data are collected with
two co-aligned narrow-field instruments: the X-ray Telescope (XRT;
Burrows \et 2004, 2005a)  and the Ultraviolet/Optical Telescope (UVOT; Roming
\et 2005).

In this paper, we report on the \swift\ observations of \grb.  The BAT
triggered and located on-board \grb\ at 2005-March-15 20:59:42 UT
(Parsons \et 2005).  The spacecraft automatically slewed to the burst
location, and the XRT and UVOT began observations starting $\approx
80$~s after the BAT trigger, one of the earliest XRT observations yet
made.  The XRT observation continued to detect the source for $\sim
10$~days, providing one of the best-sampled X-ray light curves of a
GRB afterglow to date. 
At the time of writing there are $\sim 20$ \swift\ bursts with known
redshifts, besides \grb, four of these have XRT detections out $\sim
10$ day post-burst: GRB 050319 (Cusumano \et 2005),
GRB050525a (Blustin \et 2005), GRB 050603 (Nousek \et
2005) and GRB 050401 (De Pasquale \et 2005).
The spectroscopic redshift $z=1.949$ (Kelson and Berger 2005b)
places \grb\ below the $z \approx 2.8$ mean redshift for \swift\
bursts estimated by Jakobsson \et (2005).

In this
paper we present a detailed analysis of the XRT data of \grb\ from the
first minutes to several days after the burst.
The plan for the paper is as follows.  Section~\ref{sect:obs} 
reviews the basic details of the \swift\ observations from each
instrument. Then section~\ref{sect:xrt} presents a detailed analysis
of the XRT images, light curve, and spectrum.  Due to the very high
initial count rate for the source, and the mode of the XRT camera
during the observation, the early data suffered severely from
pile-up, this problem is also discussed in section~\ref{sect:xrt},
along with a simple `workaround' solution. 
Section~\ref{sect:joint} presents a comparison of the XRT and BAT data
for the first few hundred seconds after the BAT trigger. Finally,
section~\ref{sect:disco} summarises the main results and gives a brief
discussion of some of the implications of this work.  For the purpose
of calculating luminosities the  cosmological parameters were taken to
be those of the {\it WMAP} standard cosmology, namely $ H_{0} = 70
$~km s$^{-1}$ Mpc$^{-1}$ with $ \Omega_{\rm m}=0.27,
\Omega_{\Lambda}=0.73$.

%%%%%%%%%%%%%%%%%%%%%%%%%%%%%%%%%%%%%%%%%%%%%%%%%%%%%%%%%%%%%%%%%%%%%%%%%%%%%%%

\section{Observations and basic data reduction}
\label{sect:obs}

\subsection{BAT observations}
\label{sect:obs-bat}

The BAT event data were analysed using the standard BAT analysis
software ({\tt Build 20}) as described in \swift\ BAT Ground Analysis
Software Manual (Krimm, Parsons, \& Markwardt 2004). Slew data were
processed with the corrected ray-tracing procedure for slew
data\footnote{See {\tt http:\//\//swift.gsfc.nasa.gov\//docs\//swift\\
\//analysis\//bat\_digest.html}.}  and light-curves and spectra
extracted.

\begin{figure}
\centering
\rotatebox{270}{
\epsscale{0.85}
\plotone{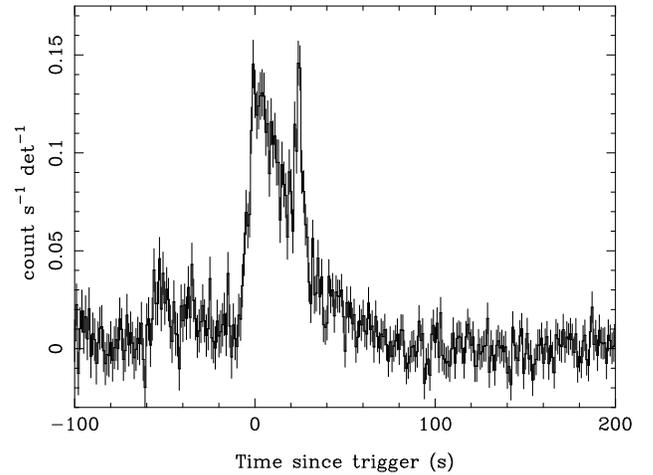}
}
\caption{ 
BAT light curve of \grb\ in $1$~s bins.
}
\label{fig:bat-lc}
\end{figure}

Figure~\ref{fig:bat-lc} shows the BAT light curve which comprises two
overlapping FRED-like peaks, 
and a possible precursor starting $\sim 60$~s before the
trigger and continuing up to the main peak. The first peak rose
over approximately $10$~s followed by a gradual decline,
interrupted by a second peak at $T_{0}+$22~s.  The burst duration
including precursor was $T_{\rm 90} = 96$ s. 
The BAT light curve from $T_0$ was binned 
such that each bin had a S:N greater than $3$ (i.e., $I/\sigma_I > 3$) and
was fitted with an exponential curve, $\exp(-t/t_e)$. Ignoring
the period $20 - 30$~s post-trigger, which was dominated by the second,
shorter peak, the exponential curve gave a good fit ($\chi^2 = 35.39$
for $32$ degrees of freedom, dof, and a rejection probability
of $p = 0.69$) with a decay constant of $t_e = 24\pm2$~s. The 
hardness ratio time series, derived from four-band BAT light curves,
clearly showed a softening of the burst spectrum with time.

The BAT spectra were extracted from the full time interval over which
the burst was detected and also intervals covering the $1$~s peak,
$T_{50}$ and $T_{90}$.  The spectra were fitted over the $15-150$~keV
range using {\tt XSPEC v11.3} (Arnaud 1996). In all cases a simple
power law provided a good fit, with no evidence for a spectral break
within the available bandpass; fitting with sharply breaking power law
or a Band function (Band \et 1993) did not substantially improve the
fit ($\Delta \chi^2 < 4$).  For the four time intervals the photon
indices ($N_{\rm ph} (E) \propto E^{-\Gamma}$) were  $\Gamma = 2.16
\pm 0.07$ (total), $2.3 \pm 0.2$ ($1$~s peak),  $2.02\pm0.07$
($T_{50}$) and  $2.13\pm 0.07$ ($T_{90}$).  The $1$~s peak flux was
$2.2\pm0.5$~ph~cm$^{-2}$~s$^{-1}$ in the $15-150$~keV band (see also
Sakomoto \et 2005; Krimm \et 2005) and the total burst fluence was
$3.4 \pm 0.3\times10^{-6}$~erg~cm$^{-2}$ (also $15-150$~keV).

If only a single photon index is measured it
is difficult to constrain the bend or peak energy of a
GRB spectrum.
In order to constrain the bend energy for \grb\ a Band function
was fitted to the BAT data from the full time interval 
assuming $\alpha = -1.3$ (the mean from the
Amati \et 2002 sample) but with all other parameters free. 
The bend energy $E_0$ was constrained to
lie below $43$~keV (in the observed frame) at the $90$\% confidence
limit (CL).
This corresponds to an upper limit on the peak energy $E_{\rm peak} = E_0
(2-\alpha)$ of $30$~keV ($90$\% CL) or $36$~keV ($99$\% CL).
Assuming $\alpha=-1.88$ (the steepest from
the Amati \et 2002 sample) gave an upper limit of $E_{\rm peak} <
31$~keV ($90$\% CL) or $40$~keV ($99$\% CL), indicating the limit on
$E_{\rm peak}$ is quite robust to the assumed value for $\alpha$.

\subsection{XRT observations}
\label{sect:obs-xrt}

At the
time of the BAT trigger the XRT was in Manual State, making
pre-planned observations of GRB 050306 in Photon Counting (PC) mode,
which meant that after the slew to \grb\ the standard set of
XRT observations were not implemented and thus the early Image Mode
(IM) snapshot, normally taken once the spacecraft has settled, was not
taken in this instance. 
See Hill \et (2004) and Burrows \et (2004) for a description of the
XRT readout modes.
The absence of IM data  immediately
following the slew prevented an early XRT position determination.
Ground analysis of the early PC data identified a new, rapidly fading
source at (J2000) RA~$= 20^{\rm h} 25^{\rm m} 54^{\rm s}$, Dec~$ = -42\degg
36 \arcmin 0.2\arcsec$ (Morris \et
2005). 

There was also $0.9$~ks of exposure taken in 
Windowed Timing (WT) mode, during orbits when the  XRT
camera was rapidly switching between PC and WT modes due to
high background. For most of these orbits there are not enough 
source counts for a robust detection and so these data were not used
in the subsequent analysis. 

The XRT data were processed by the \swift\ Data Center at NASA/Goddard
Space Flight Center (GSFC) to level $1$ data products (calibrated,
quality flagged event lists). These were further processed with the
processing pipeline {\tt xrtpipeline v0.8.8} into level $2$ data
products. The CCD operating temperature was between $-54.5$\degg C and
$-61.5$\degg C, almost $50$\degg C warmer than the original design
temperature, which led to a large number of hot and flickering
pixels. These were flagged  using the {\tt xrthotpix} tool during the
pipeline processing.  
High optical background light (e.g. due to the bright Earth limb)
dominates XRT spectra at 
low energies, these events were filtered out in the pipeline
processing and subsequently all events with energies $<0.2$~keV
were ignored.

The first useful XRT data taken during the first orbit
were $4$ frames ($10$~s) during the ``settling''  phase (when the
pointing was within $10$~arcmin but not stable), starting at $T_0 +
73.5$~s. During the first CCD frame the source is spread over the
image but it is relatively stable in the later three frames. These
frames ($7.5$~s exposure from $76-83.5$~s post-burst) 
were included in the XRT data analysis. The pointed phase PC
observation  (once the spacecraft pointing was stable) began in
earnest at $T_0+86$~s.
Following this \grb\ was observed during a further $100$ orbits of
\swift. 

%The CCD frame time in PC mode is $2.507$~s, meaning that sources
%brighter than $\gs 0.2$~ct s$^{-1}$ suffer from pile-up (Ballet
%1999). As the observed (i.e. not corrected for pile-up) count rate for
%the source was in excess of $10$~ct s$^{-1}$ during the early part of
%the first PC mode exposure, the source was severely piled-up at the
%start of the observation.  This was evident from an examination of the
%image formed from the very first few CCD frames of the pointed phase
%exposure, which showed a `hole' in the centre of the source image (see
%below).

\subsection{UVOT observations}
\label{sect:obs-uvot}

In a $100$~s exposure taken approximately $90$~s after the trigger,
UVOT detected no new source down to a $5\sigma$ limiting magnitude of
$18.5$ in V-band (Rosen \et 2005).

\subsection{Other observations}
\label{sect:obs-other}

Ground-based r-band observations with the LDSS instrument on
Magellan/Clay detected a new source within the XRT error circle
(Kelson and Berger 2005a). A $20$ min spectrum of the
afterglow identified Al~{\sc iii} $\lambda \lambda 1854.7,~1862.8$ and Si~{\sc
ii} $\lambda 1808.0$ absorption lines at a redshift $z=1.949$ (Kelson
and Berger 2005b). Using the fluence of 
$3.4\times10^{-6}$~erg~cm$^{-2}$ this implies an isotropic
equivalent $\gamma$-ray energy of $E_{\rm iso} = 3.3\times 10^{52}$~erg 
(over $15-150$~keV in the observer frame).

Soderberg and Frail (2005) reported a VLA radio counterpart at 8.5
GHz at the location of the burst. Bersier \et (2005) reported an
I-band magnitude of $20.7$, $0.48$ days after the burst trigger.
Cobb and Bailyn (2005), on behalf of the SMARTS
consortium, found an R-band decay with a slope of $-0.57$ (over
$11.6-35.6$~hr after the burst). 

%%%%%%%%%%%%%%%%%%%%%%%%%%%%%%%%%%%%%%%%%%%%%%%%%%%%%%%%%%%%%%%%%%%%%%%%%%%%%%%

\section{XRT analysis}
\label{sect:xrt}

%The X-ray afterglow of \grb\ was bright enough to cause
%substantial pile-up during the first few minutes of the
%initial PC-mode XRT exposure. 
%The final analysis of the XRT data
%began with a simple calculation, detailed below, to estimate
%the amount of pile-up as a function of the input count rate.
%With this information in hand the XRT data analysis 
%proceeded as follows. 
% The X-ray image was scrutinised
%in order to assess the significance of the pile-up and 
%a simple `workaround' solution was developed. Then the 
%X-ray light curve was examined, after dealing with the effects
%of the pile-up, and finally the X-ray spectrum was analysed.

\subsection{Pile-up estimation}
\label{sect:pile-up}

\begin{figure}
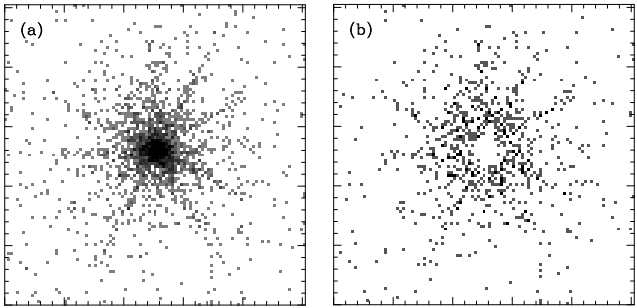

\centering
\epsscale{0.55}
\hbox{
\rotatebox{270}{
\plotone{f2a.ps}
}
\hspace{0.2 cm}
\rotatebox{270}{
\plotone{f2b.ps}
}}
\caption{ XRT images of \grb\ in the $0.2-5$~keV range shown in 
detector coordinates.
Panel (a) shows the  events accumulated from the
first $10$ orbits of pointed  exposures, the image is $100$ pixels on
a side (corresponding to $3.9$~arcmin). Panel (b) shows the image
from only the first $60$~s of exposure, when the source was
brighter than $>5$ ct s$^{-1}$, clearly showing a `hole' in the
centre of the image caused by pile-up.} 
\label{fig:xrtimg}
\end{figure}

If more than one X-ray photon is collected in a given detector
pixel in a single frame, the charges produced by the two
separate events are recorded as one. This effect is know
as `pile-up.' This is only part of the full story, however.
The charge produced by a cosmic X-ray
may be spread over one or more pixels (mono-pixel or split-pixel
events), the shape of the charge distribution determines the `grade'
of the event (or `pattern' in the \xmm\ nomenclature). Pile-up
also occurs when two X-rays are collected in neighbouring
pixels in one frame (i.e. the patterns overlap). 
Such an event might be recorded as one
split-pixel event rather than two separate events or it might be
rejected entirely as diagonal charge patterns are not produced
directly by X-rays. The effects of pile-up  are an
apparent loss of
flux, particularly from the centre of the Point Spread Function (PSF),
and a change in the grade distribution and energies of events at high input count
rates. 

%A very crude estimate of the number of piled-up events can be calculated as
%follows.  Consider a point source with a total count rate of $C$,
%imaged  such that a fraction $f$ of the counts fall on the central
%(brightest) pixel, with a CCD frame time of $\Delta T$. The mean
%number of counts per frame in the brightest pixel is then $\mu =
%C f \Delta T$.  In any given frame the actual number of counts per
%pixel, $x$, will be  drawn from a Poisson distribution with a mean of
%$\mu$, so the probability observing $x$ photons is $P(x|\mu) = \mu^{x}
%\exp(-\mu) / x!$. The probability of a frame being piled-up,
%i.e. more than one count, is $P(x>1|\mu) = 1 - P(x<2|\mu)
%= 1 - \exp(-\mu)[1+\mu]$. In order to
%restrict pile-up to the $1$ per cent level, i.e. $P(x>1|\mu) \le
%0.01$, requires that $\mu \le 0.1486$ ct pix$^{-1}$ frame$^{-1}$.  For the
%\swift\ XRT detector the frame time in PC mode is $\Delta T =
%2.507$~s, and the pixel nearest the centre of the PSF contains $f
%\approx 0.07$ of the total source counts (calculated by integrating
%the PSF model, described below, out to a $1$ pixel radius).
%These numbers imply a count rate limit of $C \le 0.85$ ct
%s$^{-1}$. Note that this is for the brightest pixel, all the other
%pixels in the source extraction region, which contain collectively the
%majority of the counts, will suffer less from pile-up, and so the
%true maximum count rate over the entire PSF will be somewhat higher than
%this estimate.

Ballet (1999) presented a very thorough treatment of flux losses
as a result of pile-up.
Equation~6 of that paper shows how the {\it observed} rate of mono-pixel
events varies with the {\it true} rate of incoming X-rays as a
function of the CCD properties and the PSF.
In order to examine at what count rates pile-up becomes significant 
for the XRT in PC mode this function was computed numerically, using
different input count rates, assuming the following instrumental parameters.
The clean (not piled-up) PSF was assumed to be a King profile (equation B1 of Ballet 1999) 
with parameters $r_c = 6.49\arcsec$ and $\beta/2 =
1.59$, as measured for the XRT at $1.49$~keV\footnote{The mean photon
  energy for \grb\ was $\approx 1.57$~keV.} from ground calibration
tests (Moretti \et 2004). 
The probability that an X-ray event produces
a CCD event with a charge pattern containing $i$ pixels was
$\alpha_i = \{ 0.778, ~0.195, ~0.014, ~0.013 \}$ for ($i=1,2,3,4$), as appropriate for
the MOS CCD at $1.49$~keV. 

The probability that an incident X-ray photon produces
a mono-pixel event is thus $\alpha_1 = 0.778$ in the limit of no pile-up.
See Ballet (1999) and Mukerjee \et (2003) for more details.
Using these numbers, and the CCD frame time of $\Delta T =
2.507$~s, the rate of mono-pixel events, with and without
pile-up, was computed as a function of input X-ray count rate.
The results are shown in Table~\ref{tab:pileup}, where
column $1$ shows the total input X-ray count rate ($\Lambda$), column $2$ shows
the expected rate of mono-pixel events assuming no pile-up ($=\alpha_1
\Lambda$), column $3$ 
shows the expected number of mono-pixel events after pile-up 
($M_1$), and column $4$ shows the ratio of mono-pixel event
count rates with and without pile-up ($\eta =M_1/\alpha_1\Lambda$).
It was evident from this calculation that even at (observed mono-pixel
event) count rates as low as $<0.1$~ct s$^{-1}$ the losses are $\sim
1$\%, and by $1$~ct s$^{-1}$ the expected flux loss is $\sim 20$ per
cent. 

\begin{table}
 \caption{Effective losses due to pile-up.}
 \begin{center}
  \begin{tabular}{llll}
\hline
Input$^{\rm a}$ & No pile-up$^{\rm b}$ & Pile-up$^{\rm c}$ & efficiency$^{\rm d}$ \\
(ct s$^{-1}$) & (ct s$^{-1}$) & (ct s$^{-1}$) &  \\
\hline
$1.29\times10^{-2}$ & $1.00\times10^{-2}$ & $9.98\times10^{-3}$ & $0.998$ \\
$1.29\times10^{-1}$ & $1.00\times10^{-1}$ & $9.78\times10^{-2}$ & $0.978$ \\
$3.21\times10^{-1}$ & $2.50\times10^{-1}$ & $2.37\times10^{-1}$ & $0.947$ \\
$6.43\times10^{-1}$ & $5.00\times10^{-1}$ & $4.49\times10^{-1}$ & $0.899$ \\
$1.28             $ & $1.00             $ & $8.18\times10^{-1}$ & $0.818$ \\
$3.21             $ & $2.50             $ & $1.63             $ & $0.652$ \\
$6.43             $ & $5.00             $ & $2.53             $ & $0.507$ \\
$1.29\times10^{1 }$ & $1.00\times10^1   $ & $3.81             $ & $0.380$ \\
$3.21\times10^{1 }$ & $2.50\times10^1   $ & $6.37             $ & $0.255$ \\
$1.29\times10^{2 }$ & $1.00\times10^2   $ & $1.30\times10^1   $ & $0.131$ \\
\hline
\end{tabular}
\end{center}

$^{\rm a}$ Rate of incoming X-rays: $\Lambda$

$^{\rm b}$ `True' rate of mono-pixel events (excluding pile-up): $\alpha_1 \Lambda$

$^{\rm c}$ `Observed' rate of mono-pixel events (including pile-up): $M_1$

$^{\rm d}$ Ratio of observed/true count rates: $\eta = M_1/\alpha_1\Lambda$

\label{tab:pileup}
\end{table}

%%%%%%%%%%%%%%%%%%%%%%%%%%%%%%%%%%%%%%%%%%%%%%%%%%%%%%%%%%%%%%%%%%%%%%%%%%%%%%%

\subsection{Image Analysis}
\label{sect:image}

Figure~\ref{fig:xrtimg} shows two images extracted from the first XRT
dataset (observation ID $00111063000$, spanning the first $10$ orbits
of data) and plotted in detector coordinates. 
Figure~\ref{fig:xrtimg}a shows the image formed from
mono-pixel ({\tt grade = 0}) events with photon energies in the
range $0.2-5$~keV accumulated during the first $10$ orbits.
There are virtually no source photons at energies $>5$~keV so 
only lower energy events were included in the analysis. Only
single-pixel events were used as these should be affected  least by
pile-up. For comparison, Figure~\ref{fig:xrtimg}b shows the image
formed from events in the same energy range, also from mono-pixel
events,  from only the first $60$~s of exposure, when the source was
brighter than $>5$ ct s$^{-1}$.
The centre of the second image is clearly deficient in counts due to
pile-up. 

\begin{figure}
\centering
\epsscale{0.85}
\plotone{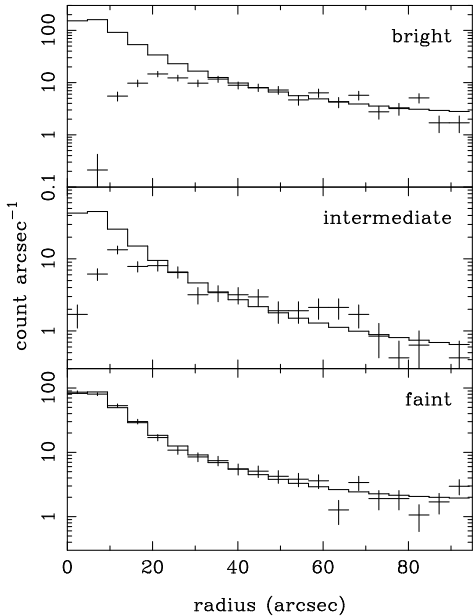}
\caption{ 
Radial profiles produced by integrating the counts in $2$-pixel
wide annuli centred on the source centroid. The three profiles were extracted
from times when the source was 
`bright' ($>5$ ct s$^{-1}$), `intermediate' ($1-5$ ct s$^{-1}$), and 
`faint' ($<1$ ct s$^{-1}$).
The data are shown with crosses, and the histogram shows the model
(PSF $+$ background) that provided a good fit to the wings of the PSF 
(see text for discussion). Clearly when the source was faint the
PSF model gave an accurate description of the radial profile, but when
the source was brighter there was a substantial loss of counts at
small radii.
}
\label{fig:psf}
\end{figure}

The effects of pile-up were clearly illustrated by an examination of
the X-ray image as a function of observed count rate. A preliminary light curve
was accumulated from single-pixel, $0.2-5.0$~keV events extracted from
a circle of radius $25$ pixels centred on the brightest pixel in the
centre of the image shown in Figure~\ref{fig:xrtimg}a. The time bin
size was set to be $25$~s (ten CCD frames).  This
light curve was used to define three time intervals: 
(i) `bright' time, when the observed source count rate was $>5$~ct s$^{-1}$;
(ii) `intermediate' time, when the count rate was $1-5$~ct s$^{-1}$; and
(iii) `faint' time, when the count rate was below $<1$~ct s$^{-1}$. 
These three time intervals covered $60$~s,
$113$~s and $12.15$~ks of exposure, respectively. For each time interval
an X-ray image was formed, and a radial profile was calculated by binning
the counts in $2$-pixel wide annuli centred on the source (taken to be
at the centroid of the faint image).

The three radial profiles were compared to a model comprising
an analytical PSF model and a constant background (per pixel)
using {\tt XSPEC v11.3}. Both the PSF and background models
were integrated in annuli to compare with the measured radial profiles:
\begin{equation}
M(r_{in},r_{out}) = \int_{r_{in}}^{r_{out}} \left[ N \left\{ 1 + \left( \frac{r}{r_c}
  \right)^2 \right\}^{-\beta/2} + B \right] 2\pi r dr
\end{equation}
where the first term in square brackets represents the King profile PSF,
with parameters $r_c$ and $\beta$,
and the second term the constant background level, $B$.
The PSF parameters were taken to be the same as used in
section~\ref{sect:spec}.

This model, with two free parameters (PSF normalisation, $N$, and
background level, $B$) was fitted to the radial profile (counts per
annulus), by adjusting the parameters to minimise the $C$-statistic
(Cash 1979), which is equivalent to finding the maximum likelihood
(ML) parameters from Poisson distributed data.
The $C$-statistic was used as the fit statistic, instead of the more
common $\chi^2$, because only the former gives the ML parameters
when there are few counts per bin, as was the case here. The
disadvantage of using the $C$-statistic is that is does
not directly provide a goodness-of-fit measure, but this
can be obtained through Monte Carlo simulations.
For each model $10^4$ simulated profiles were
generated, drawing each datum from the appropriate Poisson
distribution, and the number of simulated data with a lower
$C$-statistic (i.e. a better fit to the data) was used as a measure
of the rejection probability $p$. The analytical PSF model provided a good fit
to the faint radial profile, with $p = 0.50$, confirming this
model is indeed a good description of the source image at faint fluxes.
The measured and fitted profiles are shown in Figure~\ref{fig:psf}.

\begin{figure}
\centering
\rotatebox{270}{
\epsscale{0.85}
\plotone{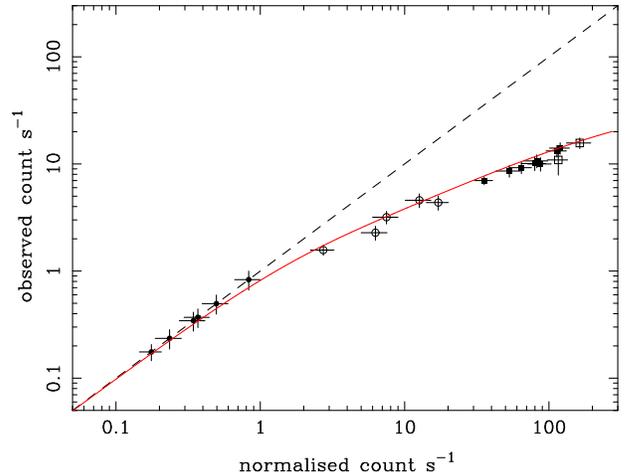}
}
\caption{ 
Count rate losses due to pile-up. 
The abscissa represents the pile-up corrected count rate
(i.e. after excluding the central part of the PSF and re-normalising
to correct for the losses), which should be comparable to the true
input source count rate. The ordinate represents the observed count
rate extracted from a simple circular extraction region (i.e. without
taking any account of pile-up). 
The hollow squares indicate the
`settling' data, the filled squares the `bright' data, 
the hollow circles the `intermediate' data and 
the filled circles the `faint data' (the latter were not
corrected for pile-up).
The dashed curve marks the expectation in the 
limit of no pile-up.
The solid curve marks the theoretically predicted relation between 
mono-pixel event count rates with and without 
pile-up effects, based on equation~6 of Ballet (1999).
}
\label{fig:corr}
\end{figure}

The same model was then fitted to the radial profiles from the
intermediate and bright images but gave
an unacceptable fit to the data, with $p > 0.9999$, in both cases,
entirely due to the loss of counts in the centre of
the image. Severe pile-up will produce a deficit of events in the
central parts of the image, but the wings of the PSF should be
relatively unaffected. Beyond some radius from the centre the
observed image should be consistent with the PSF model; this radius was  
estimated by excluding the innermost annuli from the radial profiles
until the fit became 
acceptable ($p < 0.90$). In the case of the intermediate image
($1-5$~ct s$^{-1}$),
excluding the innermost $8$ pixels radius ($19\arcsec$) gave a good fit
($p=0.797$),  while for the bright image ($>5$~ct s$^{-1}$), 
the innermost $14$ pixels ($33.0\arcsec$) had to be
excluded before the fit became acceptable ($p=0.863$). 

As a final check for the effects due to pile-up at different fluxes, 
the data were divided into finer flux intervals.
Radial profiles were extracted from times when the source
was brighter than $10$~ct s$^{-1}$ and between $0.2-1.0$~ct s$^{-1}$. 
These were fitted with the PSF model as above.
For the very brightest data, excluding the central $14$ pixels again
provided an acceptable fit to 
the data ($p=0.635$), whereas excluding only the inner
$12$ pixels did not ($p=0.969$).
These results indicate that an inner radius
cut-off of $14$ pixels (i.e. including only data from $\ge 15$ pixels
away) is sufficient to exclude the piled-up region  
of the source image even at its brightest.
Examining the image taken when the source count rate was $0.2-1.0$~ct
s$^{-1}$, the PSF model gave a good fit down to 
the innermost pixel, confirming that pile-up is a weak effect at $\ls
1$~ct s$^{-1}$ ($\ls 20$\% flux loss, see Table~\ref{tab:pileup}). 

On the basis of the above analysis, the following `workaround'
procedure was used to mitigate the adverse effects of pile-up. Source
events were extracted from  a circular region of $60\arcsec$ ($25$
pixels) radius, excluding the centre of the region when the
source was bright. In particular, for the period until $146$~s after
the burst trigger, when the observed source count rate persistently
exceeded $5$ ct s$^{-1}$, an annulus with inner and outer radii 
of $15$ and $25$ pixels was used for the extraction region.
Data from the period from $146 - 259$~s,
during which the source count rate was $1-5$ ct s$^{-1}$, were
extracted between radii of $9$ and $25$ pixels.
All data taken at
later times, when the source flux was below $\sim 1$ ct s$^{-1}$
were extracted using a full circular region.
Data extracted from annular regions were renormalised to account
for the loss of the central part of the PSF. The correction factors,
calculated by integrating the King PSF model\footnote{These correction
  factors were checked against those 
calculated using two alternative methods.
The first used the function of the encircled energy, derived from
 ground calibration data and stored in the {\tt
 swxeef20010101v001.fits} file in the \swift\ CALDB. The second folded
a $\Gamma=2$ power law spectrum through the response matrices
 generated for the appropriate source extraction regions. In all cases
the factors were very similar.}, were $14.4$ and $5.0$ for the data extracted
during bright, intermediate fluxes, respectively.

\begin{figure}
\centering
\rotatebox{270}{
\epsscale{0.85}
\plotone{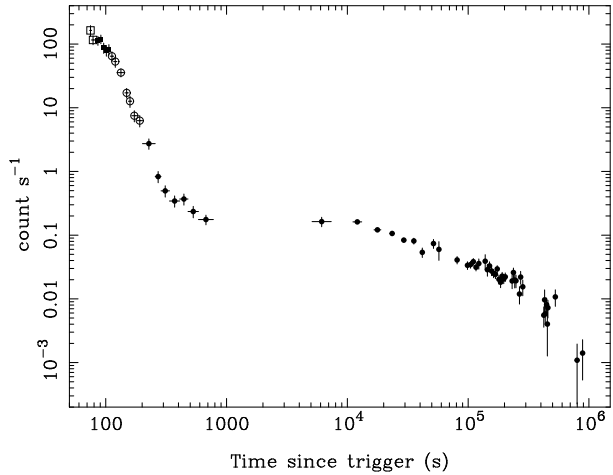}
}
\caption{ 
Light curve of \grb\ in the $0.2-5$~keV band.
These data have been corrected for pile-up; 
the different symbols reflect the different
corrections. The symbols have the same
meaning as in Fig.~\ref{fig:corr}.
Times are in the observed frame (i.e. not
corrected for cosmological time dilation).
}
\label{fig:lc1}
\end{figure}

The overall effect of pile-up can be seen by comparing the light curve
extracted from the first orbit of data before and after pile-up correction.
Figure~\ref{fig:corr} shows the relationship between the corrected and
uncorrected count rates, which indicates the effect of pile-up as a
function of source intensity. 
Also shown is the theoretical curve derived from equation~6 of
Ballet (1999), and discussed in section~\ref{sect:pile-up}, 
for the relation between mono-pixel event count rates with and without
pile-up effects. Clearly the empirical correction of pile-up 
matches the theoretical expectation based on the known CCD and PSF
characteristics.

%%%%%%%%%%%%%%%%%%%%%%%%%%%%%%%%%%%%%%%%%%%%%%%%%%%%%%%%%%%%%%%%%%%%%%%%%%%%%%%

\subsection{Timing Analysis}
\label{sect:timing}

The $0.2-5$~keV light curve of \grb\ was extracted from the full XRT
dataset. The first orbit, when the source was at its brightest, was
treated separately to avoid pile-up. The data were extracted 
in $2.507$~s bins (i.e. one CCD frame)
and then rebinned such that each time bin contained $\ge 25$ events.
(This permits the use of $\chi^2$ minimisation as a ML method.)  Error
bars were calculated using counting statistics.  The three different
time intervals during which the source was `bright,' `intermediate'
and `faint' were extracted using  the different regions, to account
for different degrees of pile-up, as discussed above.  A
background light curve was extracted, from an annulus centred on the
source with inner and outer radii of $60$ and $120$ pixels
($141-283\arcsec$),
respectively, and subtracted from the source light curve.  
The three light curves from the different
brightness intervals were renormalised to account for PSF losses;
the  resulting pile-up corrected  
light curve is shown in Figure~\ref{fig:lc1}.  Also
included in this figure are the data taken from the settling phase
(see section~\ref{sect:obs-xrt}), treated for pile-up in the same
fashion as the `bright' data.

The other orbits, for which pile-up is not an issue,  were extracted
in the standard fashion, using a $60\arcsec$ circular source region
and the same background region as for the first orbit. These data were
binned to produce one bin per orbit, accepting only those orbits
containing at least $15$ counts within the source region. The
light curve spanning all XRT observations of \grb\ is shown in
Figure~\ref{fig:lc1}.

\begin{table}
 \caption{Results of XRT light curve fitting. }
 \begin{center}
  \begin{tabular}{lccccc}
\hline
Orbits & $\alpha$               & $t_{\rm br}$ (s)      & $\alpha$            & $t_{\rm br}$ (s)              & $\alpha$                \\
\hline
$1$ & $1.8 \pm0.9$           & $115\pm12$            & $5.2_{-0.4}^{+0.5}$ & $308_{-33}^{+38}$             & $1.2\pm0.6$             \\
$2+$  & $<0.41$                &$1.2_{-0.3}^{+0.5}\times10^4$&$0.71\pm0.04$   &$2.5_{-0.3}^{+1.1}\times10^5$  &$ 2.0_{-0.3}^{+1.7}$     \\
All   & $1.9_{-1.0}^{+0.7}$    & $118\pm12$            & $5.3_{-0.4}^{+0.5}$ &                               &                         \\
      & $0.06_{-0.13}^{+0.08}$ & $1.2\pm0.4\times10^4$ & $0.71\pm0.04$       & $2.5_{-0.3}^{+1.1}\times10^5$ &  $2.0_{-0.3}^{+1.7}$     \\
\hline    
  \end{tabular}
 \end{center} 
NOTE: The models fitted to the
 first orbit ($76-778$~s post burst) and later orbits ($5\times 10^3
 - 9\times 10^5$~s) were doubly
 broken power laws, with slopes and break times as stated. The model
 fitted to all data was the sum of a singly broken power law and
 a doubly broken power law.
 \label{tab:fit}
\end{table}

The light curve was parameterised by fitting simple analytical models, comprising
connected power laws [e.g. $F(t) \propto (t-T_0)^{-\alpha}$], using {\tt XSPEC} to minimise the
$\chi^2$ fit statistic. Initially, the first orbit ($76-778$~s post
burst) and later orbits ($5\times 10^3 - 9\times 10^5$~s) were
treated separately, before fitting the entire light curve. 
Through this section, and the rest of the paper, uncertainties on 
fitted parameters correspond to $\Delta \chi^2 = 2.706$ (i.e. a
nominal $90$\% confidence region), unless stated otherwise.

\begin{figure}
\centering
\rotatebox{270}{
\epsscale{0.85}
\plotone{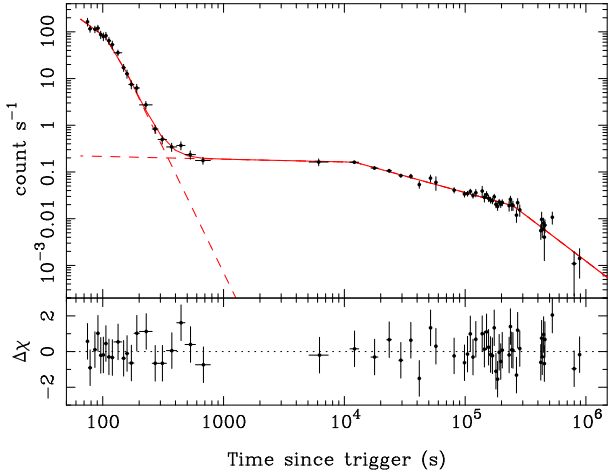}
}
\caption{ 
Light curve of \grb\ in the $0.2-5$~keV band, as in
Fig.~\ref{fig:lc1}, also showing the best-fitting model comprising
a singly broken power law and a double broken power law, which
dominate at early and late times, respectively (see
Table~\ref{tab:fit}). The lower panel
shows the $(data-model)/\sigma$ residuals.
}
\label{fig:lc2}
\end{figure}

A power law with no breaks or
just one break did not fit the first orbit data (rejection probability $p >
0.9999$), whereas a doubly broken power law gave a very good fit 
($\chi^2 = 6.90$ with $15$ dof and $p = 0.04$). 
(Even after discounting the $2$ time bins from the `settling' data
the improvement between singly-broken and doubly-broken power law
is significant at $99.4$\% confidence using an $F$-test.)
The light curve for the first $700$~s is strongly curved 
on a $\log[F_{\rm X}]-\log[t-T_0]$ plot (Figure~\ref{fig:lc1}), progressing through
flat then steep then flat phases. The best-fitting parameters for the
doubly-broken power law model are given in 
the first row of Table~\ref{tab:fit}. The initial steepening of the decay during the
first $\approx 200$~s is reminiscent of an exponential decay, as 
observed in the prompt BAT light curve
(section~\ref{sect:obs-bat}). This possibility is discussed further below.

The later orbit light curve was also inconsistent 
with a power law ($p > 0.9999$) but a singly broken power law provided a
good fit ($\chi^2 = 38.86$ with $39$ dof and $p = 0.52$). However, 
including a second break improved the fit substantially ($\chi^2 =
28.63$ with $37$ dof and $p = 0.16$). This improvement is significant
at $99.6$\% confidence, using the $F$-test. 
A smooth bend from one power law index to another gave a much worse
fit that two sharp breaks. The
late-time data therefore also show two break times (as given in the
second row of Table~\ref{tab:fit}).
Thus the complete XRT light curve for \grb\ shows at least 4 breaks if
interpreted as a series of connected power laws.
In fact, the complete light curve is well fitted by the sum of two
components, a singly broken power law dominating before $\sim T_0+300$~s
and a doubly broken power law dominating afterwards
($\chi^2 = 38.77$ for $54$ dof; $p = 0.06$). The best-fitting
parameters for this model are shown in Table~\ref{tab:fit} (rows $3$
and $4$) and the model is shown in Figure~\ref{fig:lc2}.

The only way to reconcile the early (flat-steep) part of the
light curve with a single (unbroken) power law is by allowing the
start time to be $165\pm11$~s prior to the BAT trigger, which 
lies well before the precursor in the BAT light curve (see
Figure~\ref{fig:bat-lc}), in which case the early decay slope is
$\alpha = 8.9_{-1.5}^{+0.1}$. This model gave an acceptable fit
but not as good as the the model with a break at $T_0 + 120$~s
($\chi^2 = 54.46$ for $55$ dof; $p= 0.47$).

Fitting the steep, early light curve with an exponential decay 
(plus the doubly broken power law component to fit the later
time data) also gave
a good fit, with $t_e = 35\pm2$~s, although not as good as the
broken power law ($\chi^2 = 47.76$ for $56$ dof, $p = 0.23$).
Extrapolating the exponential (prompt) plus broken power law
(afterglow) model between $10^{-2} - 10^7$~s the total
luminosity in the late-time broken power law component is $\sim 32$\%
of that in the `prompt' exponentially decaying component
(assuming no spectral evolution).

%%%%%%%%%%%%%%%%%%%%%%%%%%%%%%%%%%%%%%%%%%%%%%%%%%%%%%%%%%%%%%%%%%%%%%%%%%%%%%%

\subsection{Spectral Analysis}
\label{sect:spec}

XRT spectra were extracted from mono-pixel events collected from
source and background regions and grouped 
such that the source spectrum contained at least $20$ counts per bin. 
(Fitting the raw, un-grouped data using the $C$-statistic did not alter the
main results.) 
Five spectra were extracted from the following time intervals.
From the first pointing there were three intervals: `bright,' `intermediate' and `faint'
as discussed above.
One spectrum was extracted from the second pointing, which lasted from
$0.69-2.4 \times 10^5$~s post-burst (with an exposure time of
$40.5$~ks). This is referred to as the `mid' spectrum and lies on the
$\alpha \approx 0.7$ part of the decay light curve.
The fifth spectrum was extracted from
the fourth pointing, which lasted from
$2.7-5.3\times 10^5$~s post-burst (with an exposure time of
$18.7$~ks). This is referred to as the `late' spectrum and
lies on the $\alpha \approx 2.0$ part of the decay light curve.
(The third,  fifth and later
pointings only provided $<100$ source counts and so were not used
in the spectral analysis). The data were corrected for pile-up by
extracting source counts from annuli with radii $15-25$ and $9-25$ (inclusive)
pixels during the `bright' and `intermediate' time intervals, as for
the light curve, and using an appropriate ancillary response file to
correct for the PSF losses.

\begin{figure}
\centering
\rotatebox{270}{
\epsscale{0.85}
\plotone{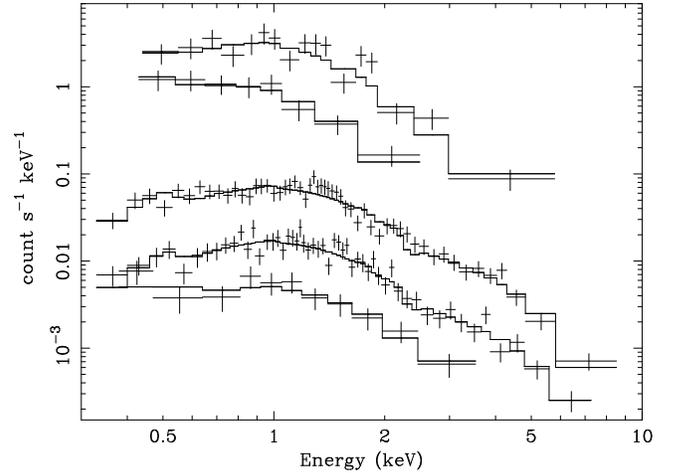}
}
\caption{ 
XRT spectra of \grb. From the top down
these correspond to the following time
intervals: $86-146$~s (`bright'), 
$146-259$~s (`intermediate'),
$259$~s to $57.39$~ks (`faint'), 
$0.69-2.4\times 10^5$~s (`mid') and
$2.7-5.3\times 10^5$~s (`late').
The data are shown with crosses and the
best-fitting absorbed power law model in each case 
(see Table~\ref{tab:fit-spec}) is shown as a histogram
}
\label{fig:fit-spec}
\end{figure}

\begin{figure}
\centering
\rotatebox{270}{
\epsscale{0.85}
\plotone{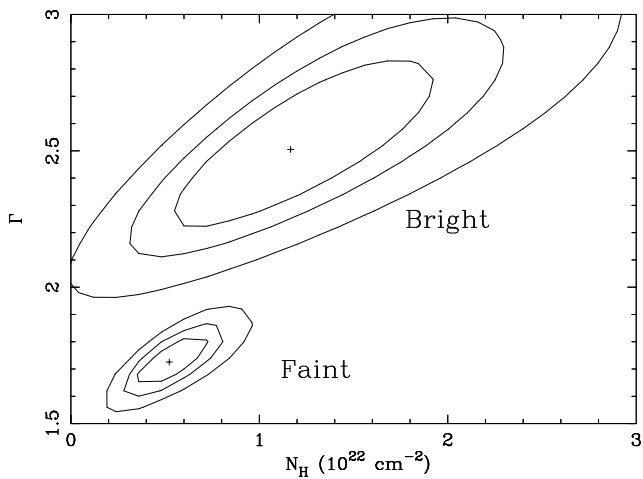}
}
\caption{ 
Confidence contours for spectral model parameters
(excess absorption column and photon index) 
as fitted to the `bright' ($86-146$~s) and `faint' ($259$~s to
$57.39$~ks)  
XRT data. The lines shown are the $\Delta \chi^ 2 = 
2.30$, $4.61$ and $9.21$ contours 
(nominal $68.3$, $90$ and $99$\% confidence bounds).
The photon index clearly decreased between `bright' and `faint'
spectra. (The contours for the `mid' and `later' data are not shown
but do overlap with those for the `faint' data.)
}
\label{fig:contour}
\end{figure}

These spectra were fitted with absorbed power law models.
The Galactic column in the direction of \grb\ is $N_{\rm H} = 4.3
\times 10^{20}$ cm$^{-2}$ (Dickey \& Lockman 1990), and this was
kept fixed, using the {\tt TBabs} model of Wilms, Allen \& McCray (2001).
A second neutral absorber in the GRB frame
(i.e. $z=1.949$) was included to model any intrinsic absorption.
The five XRT spectra are shown in Figure~\ref{fig:fit-spec}.
In all cases a power law with excess absorption provided a good
fit to the data. The results for the XRT spectral fitting are shown in 
Table~\ref{tab:fit-spec}, along with the results of fitting the BAT
spectra. There is evidence that the spectrum softened between the 
`bright' spectrum (i.e. $<T_0 + 146$~s) 
and the `faint' spectra (i.e. $>T_0 + 259$~s), but
the absorption columns remain consistent within the errors. 
This is illustrated by Figure~\ref{fig:contour} which
shows the $\Delta \chi^2$ contours
for the slope and column derived from the `bright' and `faint' data.

\begin{table}
 \caption{Results of BAT and XRT spectral fitting. }
 \begin{center}
  \begin{tabular}{lrrrrrr}
\hline
Data   & Time ($+T_0$~s) & $\Gamma$ & $N_{\rm H}$  & $\chi^2$ &  dof     & $p$  \\
\hline
BAT peak     & $25.6-26.6$  & $2.3\pm0.2$ &             & $51.79$  & $56$     &  $0.37 $\\
BAT $T_{50}$ & $0.0-24.7$   & $2.02\pm0.07$&             & $56.11$  & $56$     & $0.53 $  \\
BAT $T_{90}$ & $-48-48$     & $2.13\pm0.07$&             & $51.43$  & $56$     & $0.35 $  \\
BAT total    & $-56-69$     & $2.16\pm0.07$&             & $55.34$  & $56$     & $0.50$      \\
XRT bright   & $76-146$     & $2.5\pm0.4$  & $1.2\pm0.7$ & $21.01$  & $14$     & $0.90 $  \\
XRT inter    & $146-259$    & $2.9_{-0.8}^{+1.0}$ & $<2.1$ & $2.07$   & $5$      & $0.16 $  \\
XRT faint    & $259-5.7\times10^4$  & $1.73\pm0.11$& $0.5\pm0.2$ & $57.76$  & $51$     & $0.76 $  \\
XRT mid      & $0.69-2.4\times10^5$ & $1.79\pm0.13$& $0.8\pm0.3$ & $48.98$  & $50$     & $0.49 $  \\
XRT late     & $2.7-5.3\times10^5$  & $1.7_{-0.3}^{+0.5}$& $<1.0     $ & $5.0  $  & $8 $     & $0.24 $  \\
BAT+XRT      &              & $2.18\pm0.07      $& $0.6\pm0.4$ & $79.65$  & $71$     & $0.76 $  \\
\hline    
  \end{tabular}

NOTE: $N_{\rm H}$ is in units of $10^{22}$~cm$^{-2}$ at $z=1.949$.
 \end{center} 
 \label{tab:fit-spec}
\end{table}

%%%%%%%%%%%%%%%%%%%%%%%%%%%%%%%%%%%%%%%%%%%%%%%%%%%%%%%%%%%%%%%%%%%%%%%%%%%%%%%

\section{BAT-XRT comparison}
\label{sect:joint}

\begin{figure}
\centering
\rotatebox{270}{
\epsscale{0.75}
\plotone{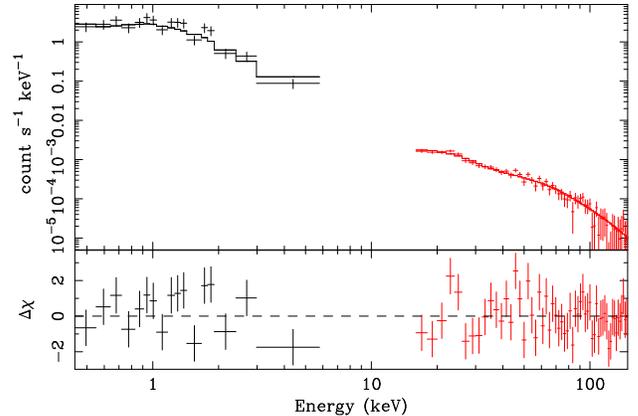}
}
\caption{ 
Spectra from the BAT (on the right) from $T_0-56$~s to $T_0+69$~s
and `bright' XRT data
(on the left) from $T_0+76$~s to $T_0+146$~s,
fitted simultaneously with a single absorbed power law ($\Gamma = 2.15
\pm 0.07$). 
}
\label{fig:joint-spec}
\end{figure}

The early (`bright') XRT spectrum (section~\ref{sect:spec}) 
showed a power law slope not dissimilar to that of the
BAT spectrum (section~\ref{sect:obs-bat}).
As a test of whether the early X-ray emission was connected to
the prompt $\gamma$-ray emission, the BAT and early XRT spectra
were fitted simultaneously. 
Figure~\ref{fig:joint-spec} shows the XRT spectrum from the
`bright' data ($\approx 86-146$~s post-trigger) and the BAT
spectrum (extracted from the full time interval) fitted
with the same absorbed power law model, but with different
normalisations between the two spectra (which allows for the temporal
decay between the times of the BAT and early XRT observations). 
A single absorbed power law gave a good fit to the 
combined data (see Table~\ref{tab:fit-spec}), consistent
with the hypothesis that the early X-ray emission and
prompt hard X-ray/$\gamma$-ray emission were produced by the same emission
spectrum. 

The spectral fits to the BAT and XRT data were used to calculate the conversion between 
observed count rate and unabsorbed $0.2-10.0$~keV flux for both BAT and XRT.
The extrapolation of the BAT spectrum into the XRT bandpass should be
reasonably accurate since, as shown above, the prompt spectrum does
extend, unbroken, into the XRT bandpass.
These factors were then used to plot the XRT and BAT light
curves in flux units, for comparison with one another. 
The error on the BAT flux resulting from the uncertainty
in the spectral model was propagated into the $0.2-10.0$~keV band
and added in quadrature with the statistical error. 
The result is shown in Figure~\ref{fig:joint-lc}.
From this figure it is clear that the tail end of the prompt emission seen
by the BAT lies very close to the early XRT data, strongly supporting
the idea that the early X-ray emission is an extension of the 
fading prompt emission.
The combined BAT-XRT light curve is dominated by an approximately 
exponential decay in flux. (The agreement between BAT and XRT light
curves does not depend on whether absorbed or unabsorbed fluxes are plotted.)

%%%%%%%%%%%%%%%%%%%%%%%%%%%%%%%%%%%%%%%%%%%%%%%%%%%%%%%%%%%%%%%%%%%%%%%%%%%%%%%

\section{Discussion}
\label{sect:disco}

\subsection{Summary of results}

The following are the main results of the XRT and BAT temporal
and spectral analyses of the \swift\ observations of \grb.

\begin{itemize}

\item
The BAT light curve showed two FRED-like peaks. The main peak
fell away exponentially with a decay constant of $\approx 24\pm2$~s in
the observer's frame ($t_e = 8\pm1$~s in the source frame). See
Figure~\ref{fig:bat-lc}. 

\item
After correcting for pile-up, the XRT light curve starting at
$T_0+76$~s showed a very rapid decline (as steep as $\alpha \gs 5$) 
until $\approx
T_0 + 300$~s. The e-folding time during this period was $35\pm2$~s
in the observer's frame ($12\pm1$~s in the source frame). 
See Figures~\ref{fig:lc1} and \ref{fig:lc2}.

\item
Despite modest spectral evolution in the BAT data, the BAT and early ($\ls T_0 +
150$~s) XRT
spectra were both consistent with steep power laws ($\Gamma \approx
2.2$). See Figure~\ref{fig:joint-spec}.

\item
Extrapolating the BAT light curve into the XRT energy band (using the
best-fitting XRT+BAT spectral model) showed the early X-ray data to 
be consistent with the tail end of the exponentially decaying prompt
emission. 
See Figure~\ref{fig:joint-lc}.

\item
The XRT spectra show significant excess absorption in the rest
frame of \grb\ ($N_{\rm H}  \sim 10^{22}$~cm$^{-2}$) with 
significant spectral hardening between early ($\ls T_0 +
150$~s) and later ($\gs T_0 + 250$~s) times.
See Figures~\ref{fig:fit-spec} and \ref{fig:contour}.

\item
After $T_0 + 300$~s the X-ray flux decays only very slowly but shows
further temporal breaks at $1.2 \pm 0.4 \times 10^4$~s 
($4.1 \pm 1 \times 10^3$~s source frame)
to a slope of
$\alpha \approx 0.7$, and again at $2.5_{-0.3}^{+1.1} \times 10^5$~s 
( $8_{-1}^{+3} \times 10^4$~s source frame)
to a steeper slope of $2.0_{-0.3}^{+1.7}$.
See Figures~\ref{fig:lc1} and \ref{fig:lc2}.

\end{itemize}

\subsection{Prompt hard X-ray/$\gamma$-rays from \grb}

\grb\ shows a relatively soft (steep) spectrum in both the
BAT and early XRT data ($\Gamma \approx 2.2$). This has
at least two interesting implications.
The first is that \grb\ may be classified as an `X-ray rich' GRB or an
`X-ray flash' (XRF). 
These classifications are often made using the softness ratio
$SR = \log [ S_{\rm X} (2-30~{\rm keV}) / S_{\gamma} (30-400~{\rm keV}) ]$
(Lamb \et 2004; Sakamoto \et 2004), with $-0.5 < SR < 0$ for an X-ray rich
burst and $SR>0$ for an XRF.
The joint BAT-XRT spectral fit (section~\ref{sect:joint}) gave
$SR = 0.2$, suggesting that \grb\ may be better classified as an XRF.

The second implication of the soft spectrum is that the energy at
which the emission peaks (the peak in $EF_{E}$ space) is below
the observed BAT energy range.
This does not match the expectations of the $E_{\rm iso} - E_{\rm
peak}$ relation discovered by Amati \et (2002).
These authors showed that the peak energy and isotropic energy 
were correlated for a small sample of GRBs with
known redshifts detected by \sax. 
Ghirlanda,  Ghisellini \& Firmani (2005)
found the same relation in BATSE data. 
Using the $E_{\rm peak} - E_{\rm iso}$ relation from Ghirlanda \et
(2005; their equation 1) the predicted peak energy for \grb\ 
is $E_{\rm peak} \sim 163$~keV in the
source frame, or $\sim 55$~keV in the observed frame, yet the observed
BAT spectrum was an unbroken power law with $\Gamma \approx 2.2$ down
to $\sim 15$~keV, suggesting a much lower $E_{\rm peak}$.
As shown in section~\ref{sect:obs-bat} the peak energy 
(in the observer's frame) was 
constrained to $E_{\rm peak} \ls 30$~keV ($90$\% CL) which is at odds
with the prediction of the Amati relation.
Furthermore, the estimated $E_{\rm iso}$ is really a lower limit 
since it was calculated over only $15-150$~keV, meaning the
predicted $E_{\rm peak}$ is a lower limit. If the true
$E_{\rm iso}$ is higher, the predicted $E_{\rm peak}$
is also higher and discrepancy with the Amati relation becomes even more
severe.  

\subsection{Prompt-afterglow transition in X-rays}

\begin{figure}[!t]
\centering
\rotatebox{270}{
\epsscale{0.85}
\plotone{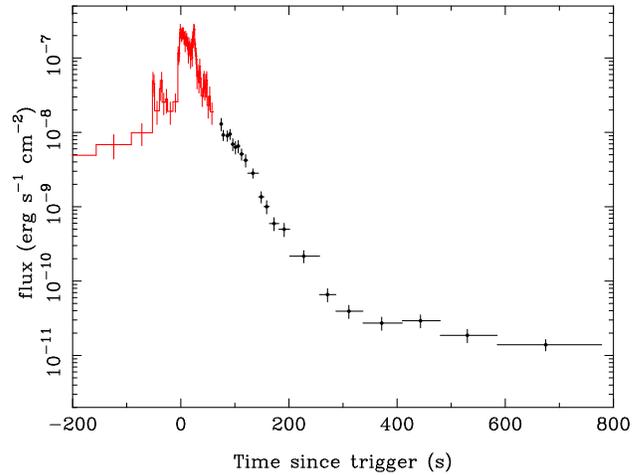}
}
\caption{ 
Light curves from the BAT (histogram) and first orbit of
XRT data (crosses) plotted in units of (unabsorbed) 
flux in the $0.2-10.0$~keV band.
The hard X-ray BAT light curve was extrapolated into the 
XRT energy band using the spectral fit discussed in 
section~\ref{sect:joint}.
}
\label{fig:joint-lc}
\end{figure}

The similarity of the spectral slopes from the early XRT data and the
prompt BAT observation (Figure~\ref{fig:joint-spec})  raises the
interesting 
possibility that the prompt hard X-rays and early soft X-rays
do not come from distinct components (`burst' and `afterglow') but are
actually different parts of the same spectrum.  
The early soft X-rays may simply
be the  lower energy emission from the same component as the prompt
hard X-ray/$\gamma$-ray emission that triggered the BAT. That the
predicted soft X-ray flux from the BAT data matches the observed flux
at the start of the XRT observation (Figure~\ref{fig:joint-lc})
strongly supports this idea; furthermore, both the prompt hard X-rays
and early soft X-rays decay in an approximately exponentially fashion
with similar e-folding timescales. Small differences in fluxes and
decay timescales between the two bands are perhaps not surprising given that
the spectrum of the prompt emission does evolve slightly (gets softer)
with time.  For example, if the X-ray flux is given by $F(t)
\propto E^{-\beta(t)}\exp(-t/t_e)$, where $\beta = \Gamma -1$, and
$\beta(t)$ increases slightly 
with time, the softer X-rays would decay slightly slower, as observed.

If the X-ray emission before $\sim T_0 + 300$~s ($100$~s in the source
frame) is dominated by the decay of the prompt burst spectrum, the
subsequent emission may be identified with the more standard X-ray
afterglow (Costa \et 1997) which must have begun as the prompt
emission decayed. The change in the X-ray spectrum after $\sim T_0 +
300$~s (see Figure~\ref{fig:contour}) supports the idea that these two
time intervals should be considered as distinct phases.  
In the following discussion 
the emission before and after $\sim T_0 + 300$~s will be referred to as
`prompt' and `afterglow,' respectively. 
In the standard relativistic
fireball models (e.g. van Paradijs, Kouveliotou \& Wijers 2000;
M{\'e}sz{\'a}ros 2002; Piran 2005) the prompt emission is caused by internal shocks within the
expanding fireball, and the afterglow is the result of the
external shocks, as the relativistic matter collides with 
circum-burst material. The X-ray to $\gamma$-ray emission observed
before $T_0+300$~s, with a spectral slope $\Gamma \approx 2.2$, is therefore
identified with internal shocks and the emission observed
afterwards, with a spectral slope $\Gamma \approx 1.8$, is 
identified with the external shocks.

Some other \swift\ bursts do not show such a strong connection
between the BAT and XRT data, e.g. GRB 050219a (Tagliaferri \et
2005). But the results for \grb\ 
are not completely without precedent. 
Tagliaferri \et (2005) also reported the \swift\ observations
of GRB 050126, with a simple FRED-like burst profile.
When extrapolated into the XRT energy range, the predicted flux at the end of
the BAT light curve is of the same order as the observed flux in the
early (from $\sim T_0 + 100$~s) XRT light curve. 
However, in this case, the prompt BAT spectrum was considerably
harder than the early XRT spectrum.
More convincing was GRB 050319 (Cusumano \et 2005; Barthelmy \et 2005) for
which the spectra from the prompt BAT and early XRT data were consistent, and 
the X-ray light curve was consistent with a single rapid decay from
prompt emission, until overtaken by a more slowly fading X-ray
afterglow at $\sim T_0 + 400$~s, which showed a harder X-ray spectrum.  
GRB 050117 (Hill \et 2005) also showed reasonable agreement between
the predicted X-ray flux from the BAT and the earliest X-ray flux
measured by the XRT, despite the complex burst profile.

Several other \swift\ XRT
observations of GRBs have revealed very rapid X-ray decays in the
first few hundred~s after the bursts (Hill \et 2005; Tagliaferri \et 2005). 
These may also be caused by the fading prompt source, perhaps off-axis
emission (Kumar \& Panaitescu 2000), in which case the mis-matches between  
BAT and XRT light curves require explanation. It is conceivable this
is due to dramatic spectral evolution. 
After the first  few $10^2$~s the X-ray emission is dominated by the
afterglow which stayed relatively constant before decaying as a broken
power law.

\subsection{The X-ray `plateau' phase}
\label{sect:plateau}

From $\approx 300$~s until
$\approx 1.2 \times 10^4$~s ($100-4\times 10^3$~s in the source frame)
the X-ray afterglow emission was almost perfectly constant ($\alpha_0 =
0.06_{-0.13}^{+0.08}$; using the convention that $F(\nu,t) \propto
t^{-\alpha} \nu^{-\beta}$, where $\beta = \Gamma -1$), beyond which it
broke to a power law decay with a slope 
of $\alpha_1 = 0.7$. Such a flat afterglow light curve is unusual and has not
been seen  in other \swift\ bursts to date
although several other bursts have shown less extreme steep-flat-steep
light curves (e.g. Chincarini \et 2005; Hill \et 2005).
It must be said however that the sampling during this period is rather
sparse and it remains possible, though perhaps unlikely, that an X-ray
re-brightening episode, such as  
observed in GRB 050406 and GRB 050502b (Burrows \et 2005b), could 
have occurred during the gaps in the light curve, masking an underlying
shallow decay. 

The indices of the temporal decay and energy spectrum of the
afterglow, $\alpha$ and $\beta$, respectively are governed by the
power law index of the energy distribution of the electrons in the
flow, $p$.  The X-ray spectrum during the plateau phase (`faint' and
`mid' spectra; Table~\ref{tab:fit-spec}) shows an energy index of
$\beta \sim 0.7$.  This rules out the X-ray band lying on the $\beta =
-1/3$ part of the synchrotron spectrum, below the emission frequency
of  lowest Lorentz factor electrons in the shock, $\nu_m$, and the
synchrotron cooling frequency, $\nu_c$ (e.g. Sari, Piran \& Narayan
1998).   However, this spectral index is close to the $\beta \sim 0.5$
expected if the X-ray band lies above $\nu_{c}$ but below $\nu_{m}$.
In this case the flux is expected to decay with $\alpha = 1/4$, which
is steeper than the observed platau phase ($\alpha \approx 0$), but a
flatter decay could occur if inverse-Compton scattering makes a
significant contribution to   electron cooling, this might be expected
for the first few hours of a  typical burst.  After $\nu_{m}$ moves
below the X-ray band (which occurs as $\nu_{m} \propto t^{-3/2}$), the
expected decay steepens to $\alpha = (3p-2)/4$, thus a slope of
$\alpha \approx 0.7$ predicts $p \approx 1.6$.  Such a low $p$ value
is unusual, but not totally without precedent (e.g., GRB 010222;
Masetti \et 2001).   Thus the break from $\alpha \approx 0$ to $0.7$,
marking the end of the plateau phase, could be due to $\nu_{m}$
moving below the X-ray band (in the standard forward shock model it is
difficult to have $\nu_{m}$ much above the XRT band after $300$~s).

Alternatively the plateau phase of the light curve may be a
consequence of `refreshed' shocks,   i.e. energy is  pumped into the
shock as it occurs.  There are two simple scenarios that will result
in refreshed shocks.  The first is that there is a distribution of
Lorentz factors  in the jet such that slower material is continuously
catching-up with faster ejecta as it decelerates in the shock
(e.g. Rees \& M{\'e}sz{\'a}ros 1998; Sari \& M{\'e}sz{\'a}ros 2000).
The second scenario occurs when the central engine stays active for a
prolonged period, and continuously injects energy into the jet at a
decreasing rate (e.g. Zhang \& M{\'e}sz{\'a}ros 2001).  After the
energy  injection ends (at $\sim12$~ks) the flux would be expected to
decay as a power law with an index $\alpha = 3(p-1)/4$ provided
that $\nu_c$ was above the XRT band. Using $p=1+2\beta \approx 2.4$
gives $\alpha \approx 1.1$ for a uniform density ISM, and steeper still for
a medium consisting of the progenitor wind, inconsistent with the
observations ($\alpha \approx 0.7$).  If, on the other hand,  $\nu_c$
was below the XRT band then $p=2\beta \approx 1.4$ and
the flux decline is given by $\alpha = -(3p-2)/4 = 0.55$ which is
closer to the observed decline.  This solution also requires
$p<2$.

\subsection{Late-time X-ray break}

The X-ray afterglow shows a second break at $2.5\times 10^5$~s
($8.4\times 10^4$~s in the source frame) to a much steeper decline,
with $\alpha_2 = 2.0_{-0.3}^{+1.7}$. This might readily be
identified with a `jet break,' corresponding to the time when
the beaming angle of the relativistic flow ($\theta_{\rm beam} \sim
\gamma^{-1}$) becomes wider than its geometrical opening angle
($\theta_{0}$), after which the  emission decays much faster (Sari,
Piran \& Halpern 1999; Rhoads 1999). 
Using equation~1 of Frail \et (2001) the observed break
timescale and $E_{\rm iso}$, yield a jet opening angle of $\theta_{0}
\sim 5$\degg, similar to the values derived for other bursts
(Frail \et 2001; Ghirlanda \et 2005).   The
predicted `true' energy released in $\gamma$-rays, corrected for
the small jet opening angle, is $E_{\gamma} = (1-\cos
\theta_{0})E_{\rm iso} \gs 1.2 \times 10^{50}$~erg, somewhat
below the mean energy found by Frail \et (2001) or Bloom \et (2003)
\footnote{\grb\ is not an
`f-GRB,' in the scheme of Bloom \et (2003), because both
the X-ray and R-band afterglow decays are very slow at $\approx
T_0+0.5$~day.}. Of course, since the prompt spectrum from the 
BAT only covers the $15-150$~keV range then the total (bolometric) 
$E_{\gamma}$ is probably a factor of a few larger, closer to the
typical values. 

If the X-rays are synchrotron emission above the
cooling frequency, $\nu_c$ then the spectrum is expected to be $\beta
= p/2$ (see Sari \et 1999; Zhang \&  M{\'e}sz{\'a}ros 2004), which
gives $p \approx 1.6$, matching the estimate above
(section~\ref{sect:plateau}) based on the temporal decay after the
plateau.  In this situation the predicted temporal slope after the jet
break is $\alpha = p \approx 1.6$ which compares reasonably with the
observed  value of $2.0_{-0.3}^{+1.7}$.  
The jet break should be achromatic and therefore the
spectrum should remain unchanged across the jet break. Indeed, within
the errors the spectral shapes  are consistent before and after this
break (the `mid' and `late' spectra of Table~\ref{tab:fit-spec}).

%%%%%%%%%%%%%%%%%%%%%%%%%%%%%%%%%%%%%%%%%%%%%%%%%%%%%%%%%%%%%%%%%%%%%%%%%%%%%%%

\acknowledgements

SV, MRG, KP, APB, OG and JPO gratefully acknowledge funding through
the PPARC, UK.  This work is supported at Pennsylvania State
University (PSU) by NASA contract NAS5-00136, at the University of
Leicester (UL) by the Particle Physics and Astronomy Research Council
on grant numbers PPA/G/S/00524 and PPA/Z/S/2003/00507, and at the
Osservatorio Astronomico di Brera (OAB) by funding from ASI on grant
number I/R/039/04. We gratefully acknowledge the contributions of
dozens of members of the XRT team at PSU, UL, OAB, GSFC, ASI Science
Data Center, and our subcontractors, who helped make this instrument
possible.
We thank an anonymous referee for a thoughtful report.

%%%%%%%%%%%%%%%%%%%%%%%%%%%%%%%%%%%%%%%%%%%%%%%%%%%%%%%%%%%%%%%%%%%%%%%%%%%%%%%

%%%%%%%%%%%%%%%%%%%%%%%%%%%%%%%%%%%%%%%%%%%%%%%%%%%%%%%%%%%%%%%%%%%%%%%%%%%%%%%

\end{document}